\documentclass[amsmath,amsfonts,amssymb,aps, prd]{revtex4}

\usepackage{hyperref}
\usepackage[english]{babel}
\usepackage{graphicx}
\usepackage{bm}
\usepackage{slashed}
\usepackage{color}
\usepackage{multirow}
\usepackage{array}
\usepackage{hhline}

\begin{document}

{
\flushright{OHSTPY-HEP-T-93-008}

\flushright{USM-TH-62}

\flushright{July 1993}

}

\title{PCAC relation for $D^\ast\to D$ axial form factors \footnote{To appear in the Proceedings of the \emph{Workshop on B Physics at Hadron Accelerators}, June 1993, Snowmass, Colorado, USA.}}

\author{Claudio O.~Dib\footnote{claudio.dib@usm.cl} }
\affiliation{Universidad T\'ecnica
Federico Santa Mar{\'{\i}}a, Valpara{\'{\i}}so, Chile}
\author{Josep Taron}
\affiliation{The Ohio State University, Columbus, OH, USA}


\preprint{OHSTPY-HEP-T-93-008
\\
USM-TH-62
\\
July 1993}
\begin{abstract}
 A relation among the form factors of $\langle D | A_\mu | D^\ast \rangle$ at $q^2 \to 0$ is derived. It is verified to lowest order in both the $1/m_c$ expansion and the expansion in number of derivatives using the effective lagrangian that incorporates the heavy quark symmetries for the $b$ and $c$ quarks, and the chiral symmetry for the $u, d, s$ quarks. 
\end{abstract}


\maketitle

The spin-flavour symmetry among hadrons containing one heavy quark ($b$, $c$) and the chiral symmetry $SU(3)_L \times SU(3)_R$ associated with the light quarks ($u$, $d$, $s$), spontaneously broken down to $SU(3)_V$ , can be invoked simultaneously to provide relations among matrix elements involving such heavy hadrons and soft Goldstone bosons ($\pi$, $K$, $\eta$). This treatment has recently been formulated under the form of an effective theory which incorporates the heavy quark and chiral symmetries at the same time [1, 2, 3, 4]. The effective lagrangian consists of an infinite number of terms, with an increasing number of derivatives, each of which is an expansion in inverse powers of the heavy quark masses. The coefficients of the different terms can not be fixed by symmetry arguments alone, and will be fitted from experiment.

To lowest order in both chiral and $1/m_Q$ expansions, the interaction lagrangian contains a term of the form (see Ref.~[5]):
\begin{equation}
-\frac{2g}{f_\pi} \left(  \partial^\nu M_{ab} D^\dagger_a D^\ast_{b\nu} + h.c.  \right), 
\end{equation}
where $M$ is the $3\times3$ Goldstone boson matrix, and $D_a$ and $D^\ast_{a\nu}$  ($a= u,d,s$) stand for the $SU(3)_V$ triplet pseudoscalar and vector fields, respectively. The constant $g$ is nothing but the effective coupling of $D^\ast D\pi$, which is a dimensionless quantity of order unity and can be fitted from $\Gamma(D^\ast\to D\pi)$. 

The object of this note is to clarify a discussion brought up by E. Levin about relations of coupling constants in the effective lagrangian. The relation derived below is the analog of the Goldberger-Treiman relation for nucleons [6], for the case of the form factors of the matrix element:

\begin{equation}
\langle D(p')| A^\mu | D^\ast (p,\epsilon) \rangle = g_1(q^2) \epsilon^\mu + g_2 (q^2) (\epsilon\cdot q) (p+p')^\mu + g_3 (q^2) (\epsilon\cdot q) q^\mu , 
\end{equation}
where $q =p-p'$, and $A^\mu$ is the axial current of the chiral symmetry, which
is conserved in the chiral limit (i.e. $m_\pi \to 0$). Thus:
\[
\langle D(p')| \partial_\mu A^\mu | D^\ast (p,\epsilon) \rangle =
(\epsilon\cdot q) \left[
 g_1(q^2) + g_2 (q^2) (m^2_{D^\ast}-m^2_D) + g_3 (q^2) q^2 \right], 
\]
\begin{equation}
= {\cal O}(m_\pi^2) \to 0 .
\end{equation}
Taking $q^2$ close to zero, the form factor $g_3(q^2)$ is dominated by the pion pole
at $q^2 = 0$. Then, for $q^2\to 0$ it follows that:
\begin{equation}
g_1(q^2) + g_2(q^2)(m^2_{D^\ast} - m^2_D) + \textrm{Res } g_3(q^2) \Big|_{q^2=0} = 0.
\end{equation}

\begin{figure}[h]
    \centering
    \includegraphics[width=0.2\textwidth]{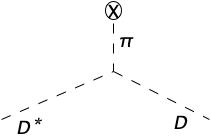}
    \caption{Diagram of the dominant term of the matrix element $\langle D(p')| A^\mu | D^\ast (p,\epsilon) \rangle$ in the limit $q^2\to 0$.}
    \label{fig1}
\end{figure}

The matrix element (2) is dominated at $q^2\to 0$  by the diagram in Fig.~1, where $\otimes$ represents the current matrix element 
$
\langle 0 | A^\mu | \pi(q) \rangle = i f_\pi q^\mu 
$,
and the $D^\ast D\pi$ vertex is $g_{D^\ast D\pi} (\epsilon\cdot q) $, where $g_{D^\ast D\pi}$ corresponds to $g$ in the effective lagrangian in Eq.~(1). One then finds
\begin{equation}
\textrm{Res } g_3(q^2)\Big|_{q^2 =0} = - f_\pi \ g_{D^\ast D\pi} .
\end{equation}
In the infinite $c$-quark mass limit, the pseudoscalar meson $D$ and the vector meson $D^\ast$ become degenerate in mass, and the relation (4) reads:

\begin{equation}
g_1(0) = f_\pi \ g_{D^\ast D\pi} .
\end{equation}

It is easy to verify that the leading terms of the lagrangian in Ref. [5] satisfy this relation for $q^\mu\to 0$. Indeed, the hadronized $A^\mu$ current reads
\begin{equation}
A^A_\mu = -2 g \left( D_a^\dagger D^\ast_b + h.c.\right) T^A_{ab} + {\cal O}(q_\mu), 
\end{equation}
which verifies Eq. (6).

Notice that relation (4) is valid for finite mass values of the heavy mesons,
and the mass difference $m_{D^\ast}-m_D$ does not originate from chiral symmetry breaking but from hyperfine splitting, which is of order ${\cal O}(1/m_Q)$.

\bigskip
\noindent
\textbf{Acknowledgements}

\medskip

We benefited from helpful conversations with J. Amundson and E. Levin. This work was supported in part by Fondecyt, Chile, grant No. 92-0806, and in part by U.S.A. grant DE-FG02-91-ER40690.

\bigskip
\noindent
\textbf{References}

\vspace{12pt}

[1] M.B. Wise, Phys. Rev. {\bf D45}, 2188 (1992).
\vspace{6pt}

[2] G. Burdman and J. Donoghue, Phys. Lett. {\bf B280}, 287 (1992).
\vspace{6pt}
 
[3] J.L. Goity, Phys. Rev. {\bf D46}, 3929 (1992). 
\vspace{6pt}

[4] T.M. Yan et al., Phys. Rev. {\bf D46}, 1148 (1992).
\vspace{6pt}

[5] M.B. Wise, Caltech preprint CALT-68-1860, 1993, unpublished. 
\vspace{6pt}

[6] M.L. Goldberger and S.D. Treiman, Phys. Rev. {\bf 110}, 1178 (1958).

\end{document}